\documentclass[journal]{IEEEtran}
\usepackage{amsmath,amsfonts}
\usepackage{algorithmic}
\usepackage{algorithm}
\usepackage{array}
\usepackage[caption=false,font=normalsize,labelfont=rm,textfont=rm]{subfig}
\usepackage{textcomp}
\usepackage{dblfloatfix}
\usepackage{url}
\usepackage{verbatim}
\usepackage{graphicx}
\graphicspath{{figures/}}
\usepackage[table]{xcolor}
\usepackage{cite}
\usepackage{booktabs} 
\usepackage{braket}
\usepackage{microtype}
\usepackage[skins]{tcolorbox}
\newtcolorbox{takeawaybox}[1][]{
  enhanced,
  width=\linewidth,
  colback=blue!5!white,
  colframe=blue!40!black,
  boxrule=0.8pt,
  arc=2.0mm,
  boxsep=0mm,
  left=1mm,
  right=1mm,
  top=1mm,
  bottom=1mm,
  fonttitle=\bfseries\normalsize,
  colbacktitle=blue!70!black,
  coltitle=white,
  title=Takeaway,
  title filled,
  toptitle=0.9mm,
  bottomtitle=0.9mm,
  lefttitle=1.0mm,
  righttitle=1.0mm,
  before skip=10pt,
  after skip=14pt,
  #1
}
\newcommand{\takeaway}[2]{\begin{takeawaybox}[title={Takeaway: #1}]#2\end{takeawaybox}}

\newtcolorbox{discussionbox}[1][]{
  enhanced,
  width=\linewidth,
  colback=red!5!white,
  colframe=red!40!black,
  boxrule=0.8pt,
  arc=2.0mm,
  boxsep=0mm,
  left=1mm,
  right=1mm,
  top=0.6mm,
  bottom=0.6mm,
  fonttitle=\bfseries\normalsize,
  colbacktitle=red!70!black,
  coltitle=white,
  title=Discussion Summary,
  title filled,
  toptitle=0.6mm,
  bottomtitle=0.6mm,
  lefttitle=1.0mm,
  righttitle=1.0mm,
  before skip=6pt,
  after skip=8pt,
  #1
}
\newcommand{\discussionsummary}[2]{\begin{discussionbox}[title={#1}]#2\end{discussionbox}}

\begin{document}

\title{A Survey of Neural Network Variational Monte Carlo from a Computing Workload Characterization Perspective}

\author{
Zhengze Xiao\textsuperscript{1,*},
Xuanzhe Ding\textsuperscript{1,*},
Yuyang Lou\textsuperscript{2},
Lixue Cheng\textsuperscript{2,\textdagger},
Chaojian Li\textsuperscript{1,\textdagger}
\\[0.5em]
\textsuperscript{1}\textit{Department of Computer Science and Engineering, Hong Kong University of Science and Technology}\\
\textsuperscript{2}\textit{Department of Chemistry, Hong Kong University of Science and Technology}
\\[0.5em]
\thanks{\textsuperscript{*}These authors contributed equally to this work.}
\thanks{\textsuperscript{\textdagger}Correspondence to: Lixue Cheng (lixuecheng@ust.hk) and Chaojian Li (chaojian@ust.hk).}
}
\maketitle

\begin{abstract}
Neural Network Variational Monte Carlo (NNVMC) has emerged as a promising paradigm for solving quantum many-body problems by combining variational Monte Carlo with expressive neural-network wave-function ans\"atze. Although NNVMC can achieve competitive accuracy with favorable asymptotic scaling, practical deployment remains limited by high runtime and memory cost on modern graphics processing units (GPUs). Compared with language and vision workloads, NNVMC execution is shaped by physics-specific stages, including Markov-Chain Monte Carlo sampling, wave-function construction, and derivative/Laplacian evaluation, which produce heterogeneous kernel behavior and nontrivial bottlenecks. This paper provides a workload-oriented survey and empirical GPU characterization of four representative ans\"atze: PauliNet, FermiNet, Psiformer, and Orbformer. Using a unified profiling protocol, we analyze model-level runtime and memory trends and kernel-level behavior through family breakdown, arithmetic intensity, roofline positioning, and hardware utilization counters. The results show that end-to-end performance is often constrained by low-intensity elementwise and data-movement kernels, while the compute/memory balance varies substantially across ans\"atze and stages. Based on these findings, we discuss algorithm--hardware co-design implications for scalable NNVMC systems, including phase-aware scheduling, memory-centric optimization, and heterogeneous acceleration.

\end{abstract}


\section{Introduction}
\label{sec:introduction}

Solving the electronic Schrödinger equation is central to quantum many-body simulation in chemistry and materials science. A long-standing challenge is the trade-off between accuracy and computational cost in traditional methods, such as coupled-cluster singles, doubles, and perturbative triples (CCSD(T)), full configuration interaction (FCI), and auxiliary-field quantum Monte Carlo (AFQMC)~\cite{hermann2023ab}. Neural Network Variational Monte Carlo (NNVMC) has emerged as a promising alternative by combining variational Monte Carlo with neural-network wavefunction modeling.

NNVMC represents the electronic wavefunction with a neural network and optimizes model parameters under the variational principle. Prior studies show that this approach can achieve strong accuracy while offering favorable asymptotic scaling in representative settings (often cited as $O(N^4)$ versus $O(N^7)$ for CCSD(T))~\cite{Carleo2017Solving}. Since Neural Quantum States were introduced, multiple neural wavefunction ans\"atze have been proposed. In this work, we focus on four representative models for ground-state problems: FermiNet~\cite{pfau2020ab}, PauliNet~\cite{hermann2020deep}, Psiformer~\cite{glehn2023a}, and Orbformer~\cite{foster2025ab}. Other models, e.g., DeepSolid~\cite{li2022ab}, LapNet~\cite{li2024computational}, QiankunNet~\cite{shang2025solving,kan2025bridging}, Neural Pfaffians~\cite{gao2024neural,gao2026excited}, and FiRE~\cite{scherbela2025accurate} are out of scope for this survey; in this study, we treat them as extensions of these parent families rather than distinct workload families.

\begin{figure*}[!t]
    \centering
    \includegraphics[width=\linewidth]{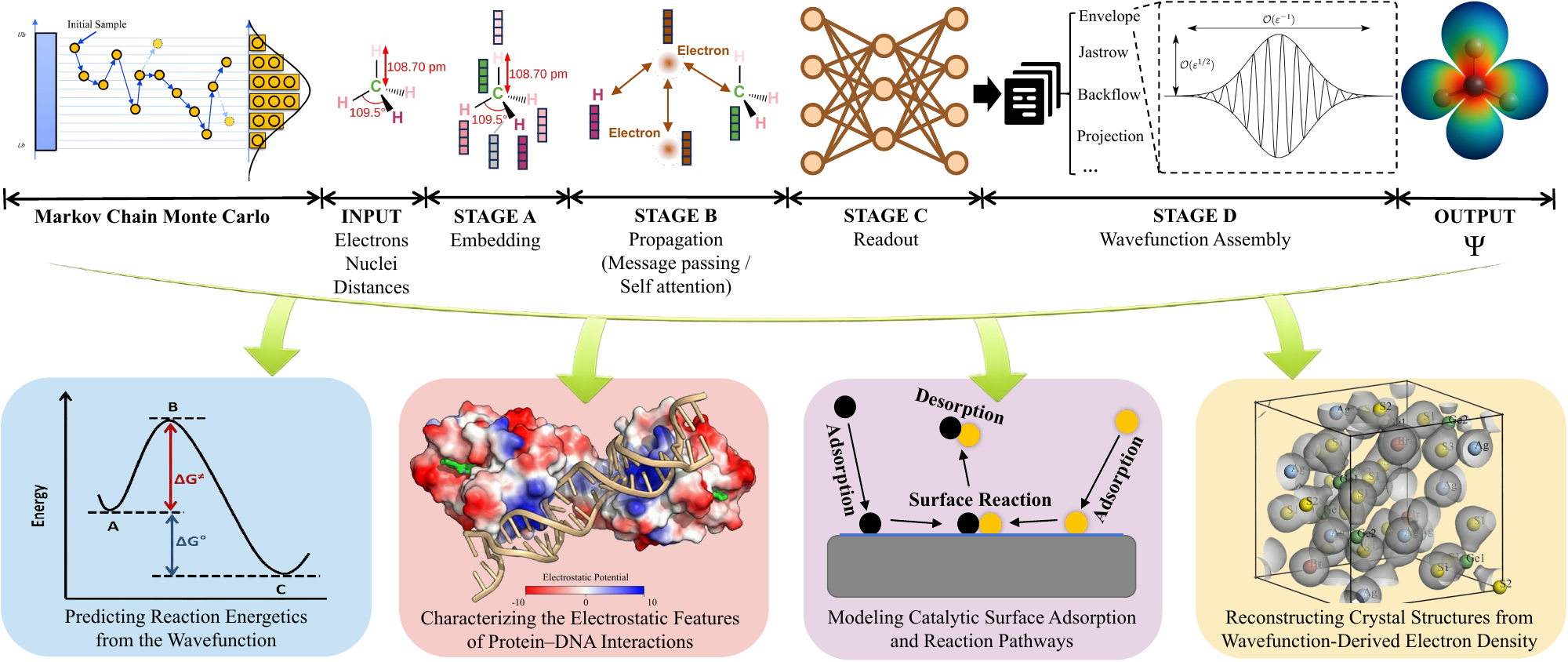}
    \caption{Overview of the end-to-end NNVMC workflow (top) and representative application domains (bottom). The workflow covers ans\"atz construction, Markov chain Monte Carlo (MCMC) sampling of electron configurations, wavefunction and local-energy evaluation, and iterative variational optimization.}
    \label{fig:overview}
\end{figure*}

Figure~\ref{fig:overview} outlines the end-to-end NNVMC workflow. Despite its algorithmic promise, practical deployment remains constrained by high computational cost~\cite{pfau2020ab,hermann2020deep,schatzle2023deepqmc,qian2025deep}. Current implementations typically handle only tens of electrons; repeated MCMC iterations require many forward evaluations~\cite{pfau2020ab,schatzle2023deepqmc}, Laplacian computation stresses memory bandwidth~\cite{pfau2020ab,schatzle2023deepqmc}, and large parameter counts increase optimization cost~\cite{hermann2020deep,glehn2023a}. These factors lead to substantial runtime and memory pressure and limit scalability to larger systems.

\par Unlike typical language or vision workloads, NNVMC has physics-specific execution stages, and kernel composition varies substantially across those stages~\cite{metropolis1953equation,hastings1970montecarlo,pfau2020ab,hermann2020deep,glehn2023a,foster2025ab,schatzle2023deepqmc}. As a result, aggregate floating-point operation (FLOP) counts alone are a weak predictor of runtime and memory behavior~\cite{schatzle2023deepqmc,qian2025deep}. This motivates dedicated stage-aware workload characterization.

Building on prior NNVMC profiling studies~\cite{schatzle2023deepqmc,qian2025deep}, this work combines a workload-oriented survey with empirical GPU characterization under a unified profiling protocol across FermiNet, PauliNet, Psiformer, and Orbformer~\cite{pfau2020ab,hermann2020deep,glehn2023a,foster2025ab}. We connect operator-level behavior to end-to-end runtime and memory trends and discuss implications for algorithm--hardware co-design in scalable NNVMC systems.

To address these challenges, this study makes the following three contributions:
\begin{itemize}
    \item We provide a workload-oriented review of representative NNVMC ans\"atze, covering PauliNet and FermiNet implemented in the \textsc{deepqmc} codebase and extending the discussion to Psiformer and Orbformer implemented in the \textsc{oneqmc} codebase~\cite{pfau2020ab,hermann2020deep,glehn2023a,foster2025ab,schatzle2023deepqmc}.
    \item We present operator- and kernel-level characterization, using empirical arithmetic intensity and roofline analysis to explain why fused elementwise and data-movement kernels often dominate runtime, even alongside general matrix multiplication (GEMM) kernels~\cite{williams2009roofline,schatzle2023deepqmc,qian2025deep}.
    \item We report hardware-level behavior, including streaming multiprocessor (SM) utilization, Tensor Core activity, memory throughput, and level-2 (L2) cache hit rate, and discuss the implications for algorithm--hardware co-design in scalable NNVMC systems.
\end{itemize}

\section{Preliminaries}
\label{sec:preliminaries}

NNVMC workloads vary widely because different ans\"atze encode different physical priors and architectural choices. They nevertheless share the same variational quantum Monte Carlo (VMC) foundation, which we review in Section~\ref{sec:prelim-vmc}. Sections~\ref{sec:prelim-paulinet-ferminet} and~\ref{sec:prelim-psiformer-orbformer} then compare how pipeline structure and dominant operator patterns differ across model families. Section~\ref{sec:prelim-ai} builds on this discussion and introduces the profiling metric used in the remainder of the paper.

\subsection{Variational quantum Monte Carlo with neural-network ans\"atze (NNVMC)}
\label{sec:prelim-vmc}

For a system with $M$ atoms and $n$ electrons, the time-independent electronic Schrödinger equation is

\begin{align}
\hat{H}\psi(\textbf{x}_1,\textbf{x}_2,...,\textbf{x}_n)=E\psi(\textbf{x}_1,\textbf{x}_2,...,\textbf{x}_n),
\end{align}
where $\textbf{x}_i =\{\textbf{r}_i, \sigma_i\}$, with $\textbf{r}_i$ and $\sigma_i$ denoting the position and spin of electron $i$. Here, $E$ is the energy eigenvalue, $\psi$ is the wavefunction, and $\hat H$ is the Hamiltonian operator. Under the Born--Oppenheimer approximation, the nuclei are treated as classical particles with fixed positions, and $\hat H$ becomes

\begin{align}
\hat{H}=&-\frac{1}{2}\sum_i\nabla_i^2 + \sum_{i}\sum_{j<i}\frac{1}{|\mathbf{r}_i-\mathbf{r}_j|} \nonumber \\ 
&- \sum_i\sum_I\frac{Z_I}{|\mathbf{r}_i-\mathbf{R_{I}}|} + \sum_{I}\sum_{J<I}\frac{Z_{I}Z_{J}}{|\mathbf{R}_I-\mathbf{R}_J|},
\end{align}
where $Z_I$ and $\textbf{R}_I$ are the charge and position of nucleus $I$, and $\nabla_i^2$ is the Laplacian with respect to electron coordinate $\mathbf{r}_i$.

Classical quantum chemistry typically relies on hand-designed ans\"atze. NNVMC instead parameterizes the wavefunction with a neural network $\psi_\theta$ and optimizes the parameter set $\theta$ under the variational principle. In this paper, we focus on ground-state problems, for which the target parameter set $\theta'$ minimizes the expected Hamiltonian:
\[
\mathbf{\theta'}= \arg \min_{\theta} \langle \hat{H} \rangle_{\psi_{\theta}}.
\]
The objective is estimated by sampling $n_\text{samples}=n_b \times n_s$ electron configurations $\{r\}$ from $|\psi_{\theta}|^2$, where $n_b$ denotes the number of parallel walkers and $n_s$ denotes the number of sampling steps, and then evaluating $\psi_\theta(\{r\})$:

\begin{align}
\mathcal{L}(\theta) &= \langle{\hat{H}}\rangle_{\psi_\theta}  =\frac{\langle{\psi_\theta|\hat{H}|\psi_\theta}\rangle}{\langle{\psi_\theta|\psi_\theta}\rangle} \nonumber \\ &= \mathbb{E}_{\{r\} \sim |\psi_{\theta}|^2} E_{\mathrm{loc}}[\psi_{\theta}],
\label{eq:loss}
\end{align}
where the local energy is
\begin{align}
E_{\mathrm{loc}}[\psi_{\theta}] = T_{\mathrm{loc}}(\{r\}) + V(\{r\}),
\label{eq:eloc}
\end{align}
where $T_{\mathrm{loc}}$ denotes the local kinetic energy and $V$ is the Coulomb potential. These two terms are given by
\begin{align}
T_{\mathrm{loc}}(\{r\}) = -\frac{1}{2}\sum_i\frac{\nabla_{\mathbf{r}_i}^2\psi_{\theta}(\{r\})}{\psi_{\theta}(\{r\})},
\label{eq:kinetic} 
\end{align}
\begin{align}
V(\{r\}) &= \sum_{i<j}\frac{1}{|\mathbf{r}_i - \mathbf{r}_j|} - \sum_{i,I}\frac{Z_I}{|\mathbf{r}_i - \mathbf{R}_I|} + \sum_{I<J}\frac{Z_I Z_J}{|\mathbf{R}_I - \mathbf{R}_J|}.
\label{eq:potential}
\end{align}
In practice, optimization uses the following unbiased estimator of the loss gradient:
\begin{align}
\begin{split}
\nabla_\theta\mathcal{L}(\theta)
&= 2\mathbb{E}_{\{r\} \sim |\psi_{\theta}|^2}\Big[
(E_{\mathrm{loc}}[\psi_{\theta}](\{r\})-\mathcal{L}(\theta)) \\
&\qquad\qquad\qquad\cdot \nabla_\theta \log|\psi_{\theta}(\{r\})|
\Big]
\end{split}
\end{align}

Optimization yields the converged ground-state wavefunction $\psi_{\theta'}$ and the corresponding expected energy $E_0$:
\begin{align}
    \braket{E_0} = \frac{1}{n_b n_s}\sum_{i=1}^{n_b}\sum_{j=1}^{n_s}E_{\mathrm{loc}}[\psi_{\theta'}](\{r\}_{ij})
\end{align}

\noindent\textbf{Highlight 1 (Optimization objective).}
NNVMC directly parameterizes the many-electron wavefunction with a neural network and optimizes it through stochastic estimation of Eq.~(\ref{eq:loss}) and its gradient. This unifies physical energy minimization and model training in one variational loop.

\noindent\textbf{Highlight 2 (Execution structure).}
Each optimization step jointly performs MCMC sampling, wavefunction evaluation, and derivative/Laplacian computation. This coupling is a key reason NNVMC runtime behavior differs from standard deep-learning workloads and motivates stage-aware profiling in later sections.
Figure~\ref{fig:overview} summarizes this end-to-end workflow.

Beyond the released \textsc{oneqmc} models profiled here, other potentially more efficient model designs may also exist. For example, approximating the Laplacian through Hutchinson trace estimation (HTE) is a potential direction for improving the efficiency of local-energy evaluation~\cite{hu2024hte}. However, since no corresponding public model release is currently available, we leave profiling of such designs to future work.

Sections~\ref{sec:prelim-paulinet-ferminet} and~\ref{sec:prelim-psiformer-orbformer} then turn to model-specific architecture details and execution pipelines for PauliNet, FermiNet, Psiformer, and Orbformer in the \textsc{deepqmc} and \textsc{oneqmc} codebases.

\begin{figure*}[!t]
    \centering
    \includegraphics[width=\linewidth]{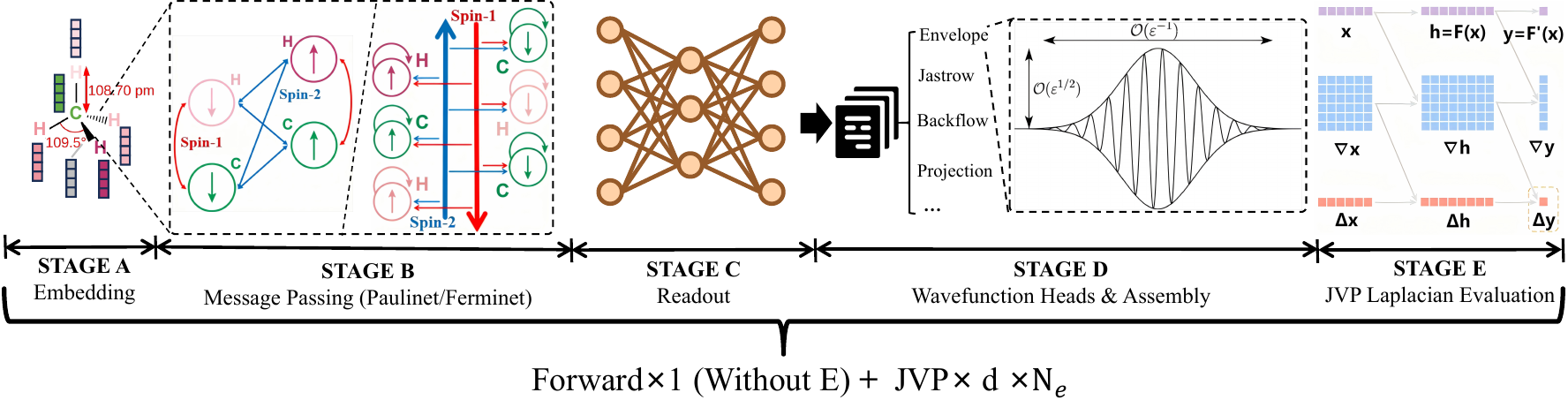}
    \caption{Overview of the execution pipeline of PauliNet and FermiNet in the \textsc{deepqmc} codebase. Stages~A--E represent feature construction and embedding, electron-correlation updates through neural blocks, readout projection, Slater-determinant wavefunction assembly, and derivative/Laplacian evaluation for local-energy computation. The bracket below the pipeline indicates that one VMC evaluation includes one forward pass through Stages~A--D and then a Stage~E replay over all Cartesian directions of all electrons through JVP, summarized in the figure as a cost scaling with $d \times N_{\mathrm{e}}$, where $d$ is the spatial dimensionality and $N_{\mathrm{e}}$ is the number of electrons. The rightmost inset schematically illustrates the JVP-based Laplacian evaluation in Stage~E: it traces an input quantity $x$, an intermediate state $h=F(x)$, and an output $y=F'(x)$, while the paired $\nabla(\cdot)$ and $\Delta(\cdot)$ annotations indicate the gradient- and Laplacian-related quantities propagated in this procedure.}
    \label{fig:paulinet-pipeline}
\end{figure*}

\subsection{PauliNet and FermiNet via the \texttt{deepqmc} codebase}
\label{sec:prelim-paulinet-ferminet}

In this work, PauliNet and FermiNet are treated as determinant-based NNVMC ans\"atze. We use \textsc{deepqmc} for implementation and profiling~\cite{hermann2020deep,pfau2020ab,schatzle2023deepqmc}. Figure~\ref{fig:paulinet-pipeline} summarizes their stage-wise execution flow.

\noindent\textbf{Pipeline overview.}
Each VMC iteration samples electron configurations from $|\psi_\theta|^2$ using Markov chain Monte Carlo, typically Metropolis--Hastings, and then evaluates wavefunction- and local-energy-related quantities for those samples~\cite{metropolis1953equation,hastings1970montecarlo,schatzle2023deepqmc}. The neural-network forward path is therefore embedded in the sampling loop and executed repeatedly.

\begin{itemize}
    \item \textbf{Stage A (feature construction).} Electron coordinates are transformed into physically motivated inputs, including electron--nuclear and electron--electron distances, relative coordinate vectors, and spin features.
    \item \textbf{Stage B (correlation update).} Permutation-equivariant neural layers update electron representations to capture many-body correlations and provide the structural basis for fermionic antisymmetry.
    \item \textbf{Stage C (readout projection).} Updated features are projected into intermediate orbital-like representations for wavefunction construction.
    \item \textbf{Stage D (wavefunction assembly).} Multiple Slater determinants are formed and combined to produce $\psi_\theta(\mathbf{R})$, optionally with additional correlation terms such as Jastrow/backflow components~\cite{pfau2020ab,lopezrios2006inhomogeneous}. In practice, outputs are commonly represented by logarithmic amplitude and sign for numerical stability.
    \item \textbf{Stage E (derivative/Laplacian evaluation).} Local-energy evaluation requires first- and second-order derivatives. In \textsc{deepqmc}, this stage uses Jacobian--vector product (JVP)-based replay of earlier computations to obtain gradient and Laplacian terms~\cite{baydin2018automaticdifferentiation,schatzle2023deepqmc}.
\end{itemize}

\noindent\textbf{Summary.}
For PauliNet/FermiNet in \textsc{deepqmc}, a single evaluation consists of Stages~A--D followed by derivative evaluation in Stage~E, all within the MCMC loop. This execution pattern is the basis for the workload characterization in later sections.

\begin{figure}[!b]
    \centering
    \includegraphics[width=\linewidth]{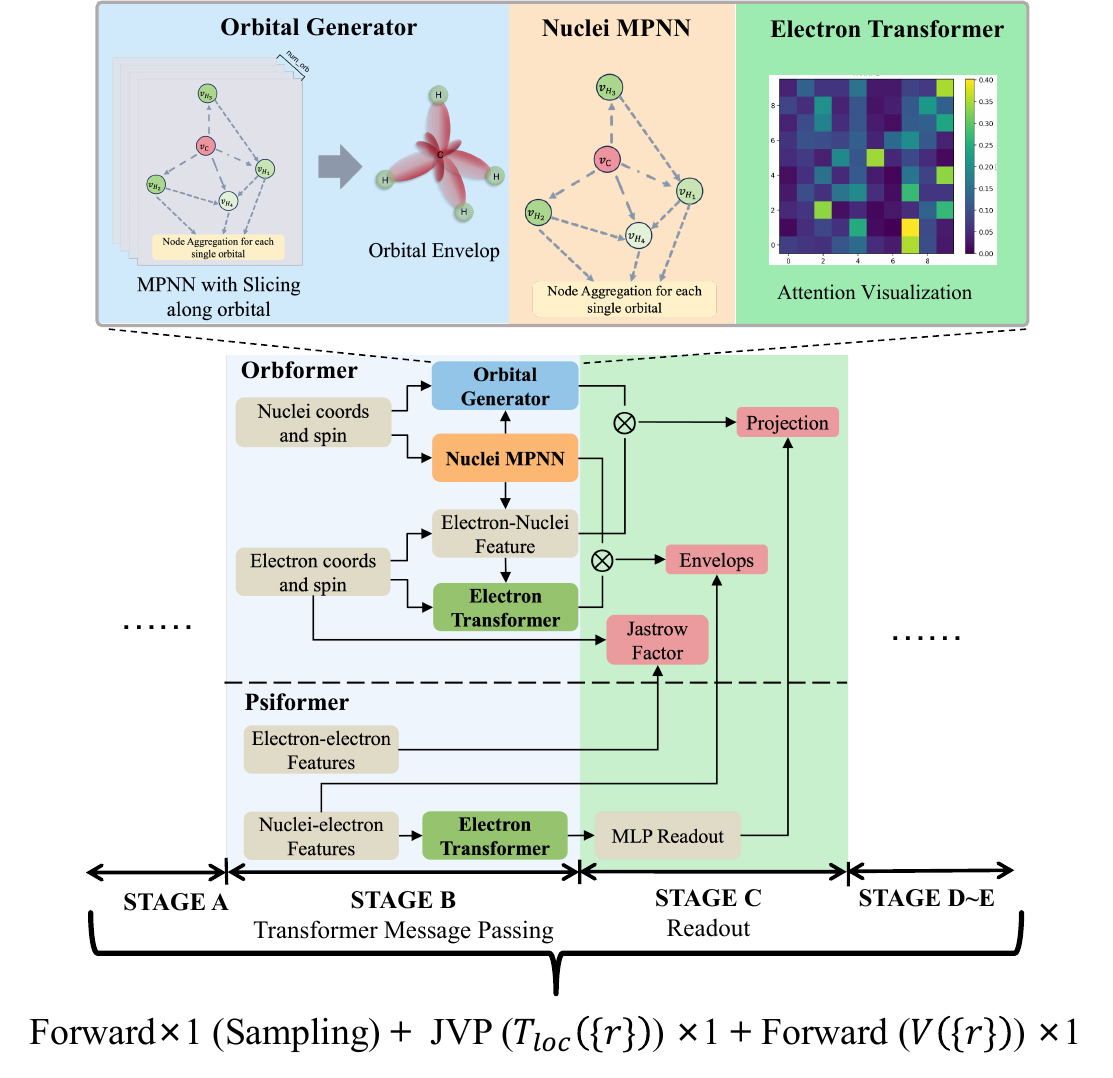}
    \caption{Overview of the execution pipeline of Psiformer and Orbformer in the \textsc{oneqmc} codebase. The upper panels show the Orbformer-specific modules used in Stage~B, namely the orbital generator, nuclei message-passing neural network (MPNN), and electron Transformer, while the lower block diagram contrasts the Orbformer path with the simpler Psiformer path. Specifically, Stage~B (Transformer-based electron representation learning) and Stage~C (orbital/readout projection) are the main departures from \textsc{deepqmc}; Stages~A and D (feature construction and Slater-determinant wavefunction assembly) follow the same high-level structure as in Figure~\ref{fig:paulinet-pipeline}, while Stage~E uses a forward Laplacian estimator and is omitted here for brevity.}
    \label{fig:orbformer-pipeline}
\end{figure}

\subsection{Psiformer and Orbformer via the \texttt{oneqmc} codebase}
\label{sec:prelim-psiformer-orbformer}

Within the variational quantum Monte Carlo (VMC) framework, Psiformer and Orbformer~\cite{glehn2023a, foster2025ab} are implemented in the \textsc{oneqmc} codebase. Figure~\ref{fig:orbformer-pipeline} summarizes the corresponding execution flow.

\noindent\textbf{Pipeline overview.}
As in \textsc{deepqmc}, each VMC iteration samples configurations from $|\psi_\theta|^2$, evaluates wavefunction-related quantities, and computes derivatives for local-energy estimation. The forward path remains embedded in the Monte Carlo loop and is executed repeatedly.

\begin{itemize}
    \item \textbf{Stage A (feature construction).} Atomic/system coordinates are transformed into electron- and nucleus-related input features; Orbformer additionally includes nucleus-side features for its Nuclei MPNN~\cite{foster2025ab}.
    \item \textbf{Stage B (Transformer interaction).} Psiformer and Orbformer use a single-stream ElectronTransformer with multi-head self-attention for electron-representation updates~\cite{glehn2023a,foster2025ab}. Orbformer further introduces an Orbital Generator and a Nuclei MPNN to support cross-system transferability.
    \item \textbf{Stage C (readout projection).} Electron representations are projected into orbital/readout quantities used for determinant construction (multi-layer perceptron (MLP) projection in Psiformer; Orbital-Generator-conditioned projection in Orbformer).
    \item \textbf{Stage D (wavefunction assembly).} Slater determinants are assembled from Stage~C outputs and combined with Jastrow terms to produce $\psi_\theta(\mathbf{R})$~\cite{glehn2023a,foster2025ab}. In practice, outputs are commonly represented in logarithmic form for numerical stability.
    \item \textbf{Stage E (derivative/Laplacian evaluation).} Local-energy evaluation includes a kinetic-energy (KE) term $T_{\mathrm{loc}}(\{r\})$ and a potential-energy (PE) term $V(\{r\})$, as shown in Eq.~(\ref{eq:eloc}). Compared with \textsc{deepqmc}, \textsc{oneqmc} uses a 
    forward Laplacian~\cite{li2024computational} estimator for the KE term,
    improving the Laplacian computation by directly propagating first-order derivatives and Laplacian terms through the network in a single forward pass without any explicit Hessian constructions; the PE term is evaluated directly from the sampled coordinates. This keeps the number of differentiation passes constant rather than scaling with all Cartesian directions of all electrons~\cite{baydin2018automaticdifferentiation}.
\end{itemize}

\noindent\textbf{Summary.}
For Psiformer/Orbformer in \textsc{oneqmc}, Stages~A--D provide forward wavefunction evaluation and Stage~E provides local-energy evaluation through a forward Laplacian path, again within the MCMC loop. This execution pattern supports the profiling analysis in Section~\ref{sec:profiling}.

\subsection{Arithmetic Intensity as a Performance Metric}
\label{sec:prelim-ai}
Our performance analysis operates at GPU-kernel granularity. Each kernel corresponds to a numerical operator, such as dense matrix multiplication, fused elementwise transformation, or Jacobian--vector product computation. We summarize each kernel by its arithmetic intensity (AI), defined as the ratio between floating-point operations (FLOPs) and off-chip memory traffic:
\begin{equation}
    \mathrm{AI} = \frac{\text{FLOPs}}{\text{Bytes moved}}.
\end{equation}
Kernels with high AI perform many floating-point operations per byte transferred and are therefore more likely to approach compute limits, whereas low-AI kernels are typically bandwidth-limited. In the roofline model, AI determines whether a kernel lies in the memory-bound or compute-bound region and thus provides a compact way to interpret NNVMC behavior on GPUs.

We use empirical AI reported by NVIDIA Nsight Compute rather than hand-derived FLOP formulas. For each kernel instance, Nsight Compute provides hardware-counter estimates of executed floating-point operations and off-chip memory traffic; Section~\ref{sec:profiling} aggregates these measurements by kernel family and model stage. This empirical view is especially useful for heterogeneous NNVMC workloads: GEMM kernels in embedding and message-passing blocks often show relatively high AI, whereas fused elementwise kernels typically remain at very low AI. In the \textsc{deepqmc} PauliNet/FermiNet path, JVP-based replay during Laplacian evaluation further introduces long sequences of fine-grained, bandwidth-sensitive kernels. These AI patterns help explain the roofline results and why end-to-end execution can remain memory-bound despite the presence of large matrix multiplications.

\section{Profiling}
\label{sec:profiling}
\begin{figure*}[!t]
    \centering
    \includegraphics[width=0.9\linewidth]{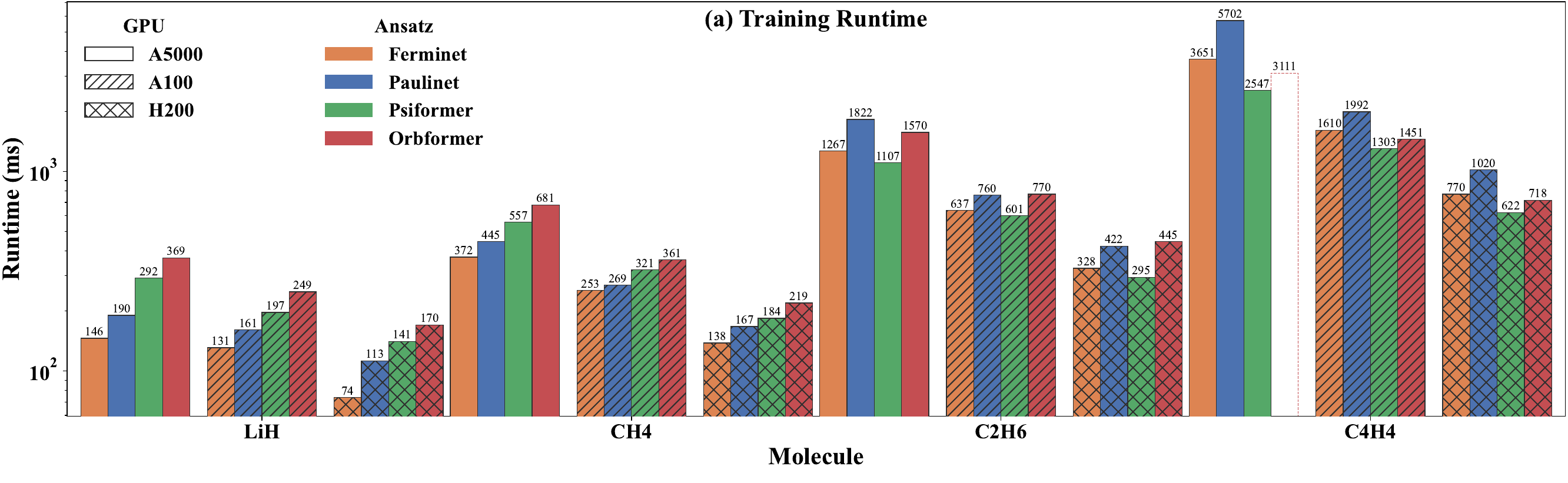}\\[-0.2em]
    \includegraphics[width=0.9\linewidth]{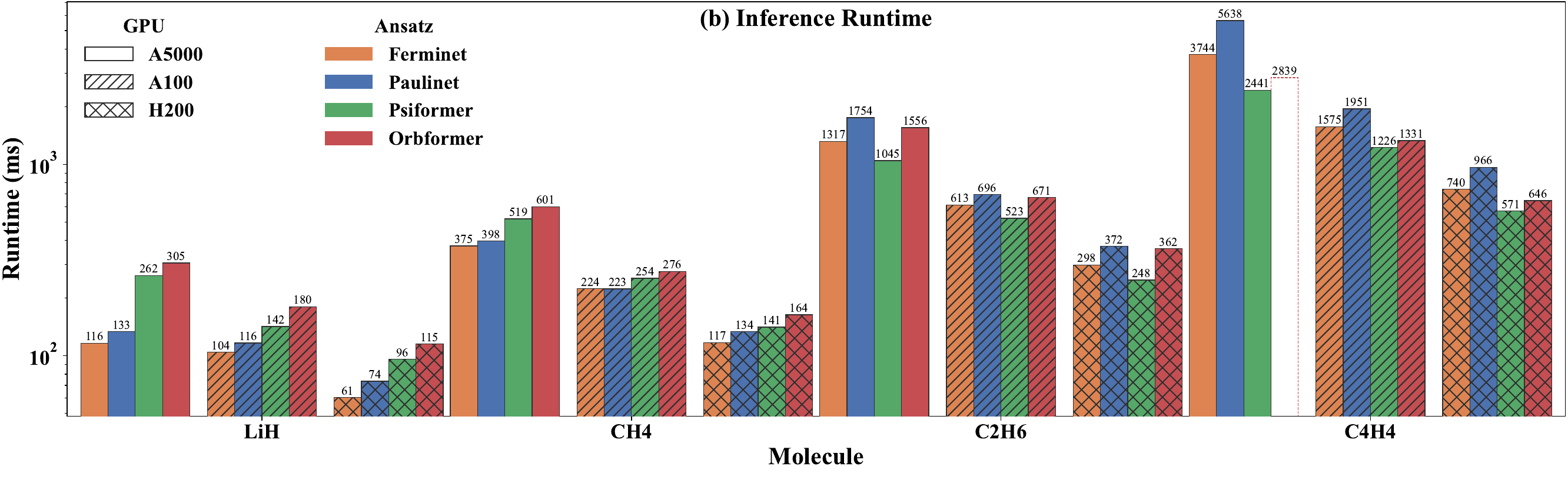}\\[-0.2em]
    \includegraphics[width=0.9\linewidth]{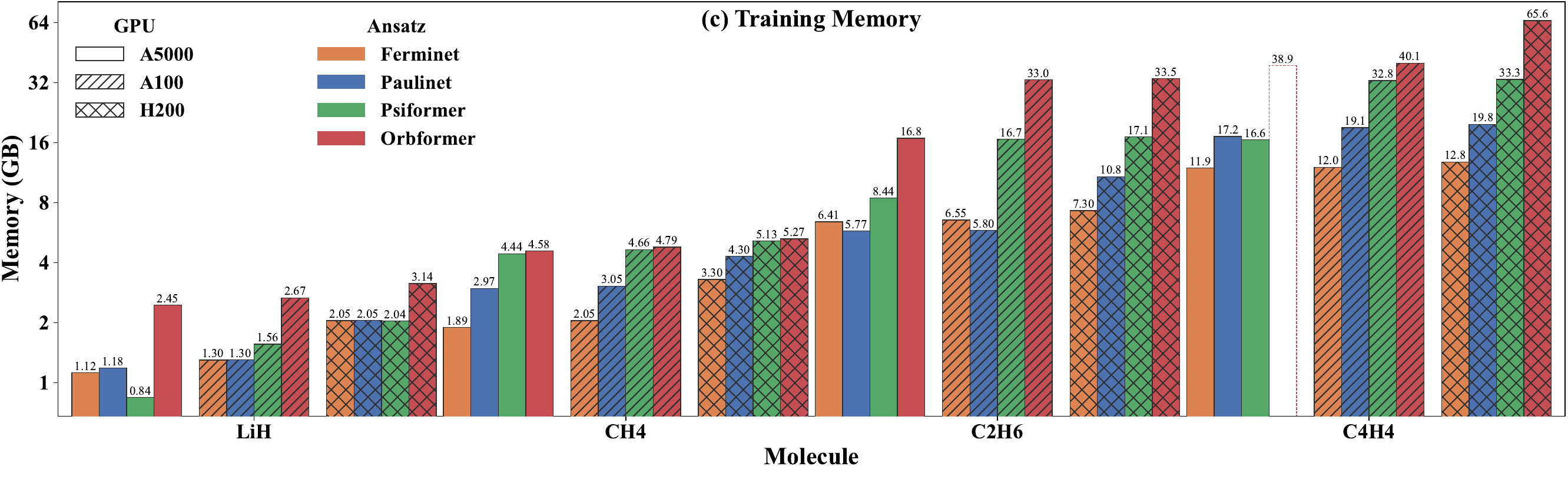}\\[-0.2em]
    \includegraphics[width=0.9\linewidth]{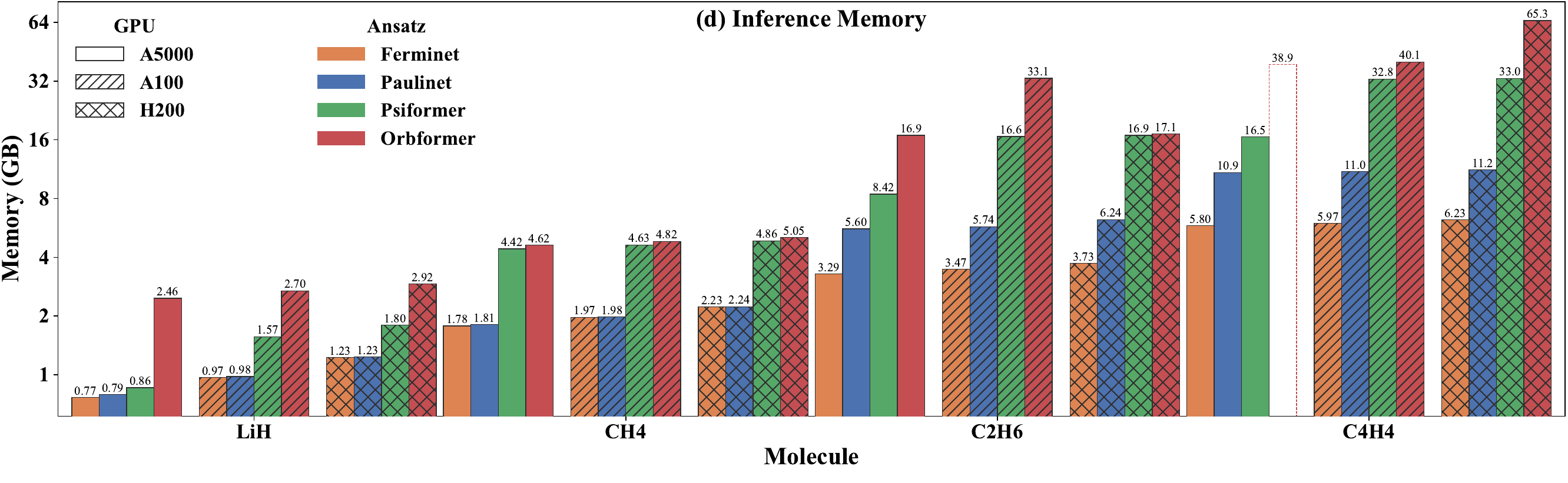}
    \vspace{-1em}
    \caption{Comparison of GPU runtime and memory usage across four molecules, four determinant-based NNVMC ans\"atze, and three GPUs (A5000, A100, and H200). From top to bottom, the panels report training runtime, inference runtime, training memory, and inference memory. Runtime uses log scale and memory uses log2 scale. Training runtime denotes one optimization step (forward, backward, and optimizer update), while inference runtime includes MCMC sampling and local-energy evaluation with model-specific Stage~E Laplacian implementations. Dashed outlines indicate that Orbformer encounters out-of-memory issues on the A5000 for C$_2$H$_6$ and C$_4$H$_4$.}
    \label{fig:runtime-log}
\end{figure*}

We profile determinant-based NNVMC workloads at two complementary levels: end-to-end runtime/memory and kernel-level execution behavior~\cite{schatzle2023deepqmc,glehn2023a,foster2025ab}. We first report system-level trends across molecules and GPUs using Figure~\ref{fig:runtime-log}, and then analyze dominant kernel families and bottlenecks.

\subsection{Profiling Settings}
\label{sec:profile-settings}
We summarize the experimental setup by category:
\begin{itemize}
    \item \textbf{Hardware platform.} Kernel-level profiling is conducted in inference mode on an NVIDIA RTX~A5000 GPU (24~GB).
    \item \textbf{Model implementations and software stack.} For PauliNet and FermiNet, we use JAX~0.4.28 (jaxlib~0.4.28)~\cite{jax2018github} with \textsc{deepqmc}~1.2.0~\cite{schatzle2023deepqmc} on CUDA~12.4. Psiformer and Orbformer are profiled in the \textsc{oneqmc} codebase under the same software and platform configuration.
    \item \textbf{Profiling tools.} Execution traces are collected with Nsight Systems~2025.5.1~\cite{nvidiaNsightSystems}, and kernel-level metrics are collected with Nsight Compute~2025.3.1~\cite{nvidiaNsightCompute}.
    \item \textbf{Timing and statistical protocol.} We report GPU kernel time only: CUDA kernel durations are summed, while host-side overheads, memory-copy operations, and just-in-time (JIT) compilation time are excluded. Each workload is executed 11 times; the first run is treated as warm-up and discarded, and we report the mean of the remaining 10 runs.
    \item \textbf{Kernel-family attribution.} Kernels are grouped into families using name-based regular expressions. Since XLA (Accelerated Linear Algebra) JIT fusion in the JAX stack~\cite{jax2018github,xla2026github} can combine elementwise operations, layout transforms, and small GEMM-like work into one kernel, family attribution is approximate.
    \item \textbf{Workload configuration.} We use 32-bit floating point (FP32) and a batch of 1024 electron configurations (parallel MCMC walkers) for all experiments in this section. For kernel-level profiling, we focus on methane (CH$_4$; 10 electrons). For PauliNet and FermiNet, each profiling run uses default \textsc{deepqmc} sampling settings and consists of ten Metropolis--Hastings MCMC transitions~\cite{metropolis1953equation,hastings1970montecarlo}, one forward pass through Stages~A--D, and $3N_{\mathrm{e}}$ JVP replays for Laplacian evaluation (Stage~E), which re-execute computations from Stages~A--D. For Psiformer and Orbformer, we use corresponding \textsc{oneqmc} defaults with the same precision and batch settings; each run contains one forward pass through Stages~A--D and one Stage~E local-energy path based on a forward Laplacian estimator~\cite{li2024computational}, within the same Metropolis--Hastings process.
\end{itemize}
Unless otherwise stated, all analyses in Sections~\ref{sec:profile-paulinet}--\ref{sec:profile-orbformer} use the profiling setup defined in this subsection.

\begin{figure*}[!b]
    \centering
    \includegraphics[width=\linewidth]{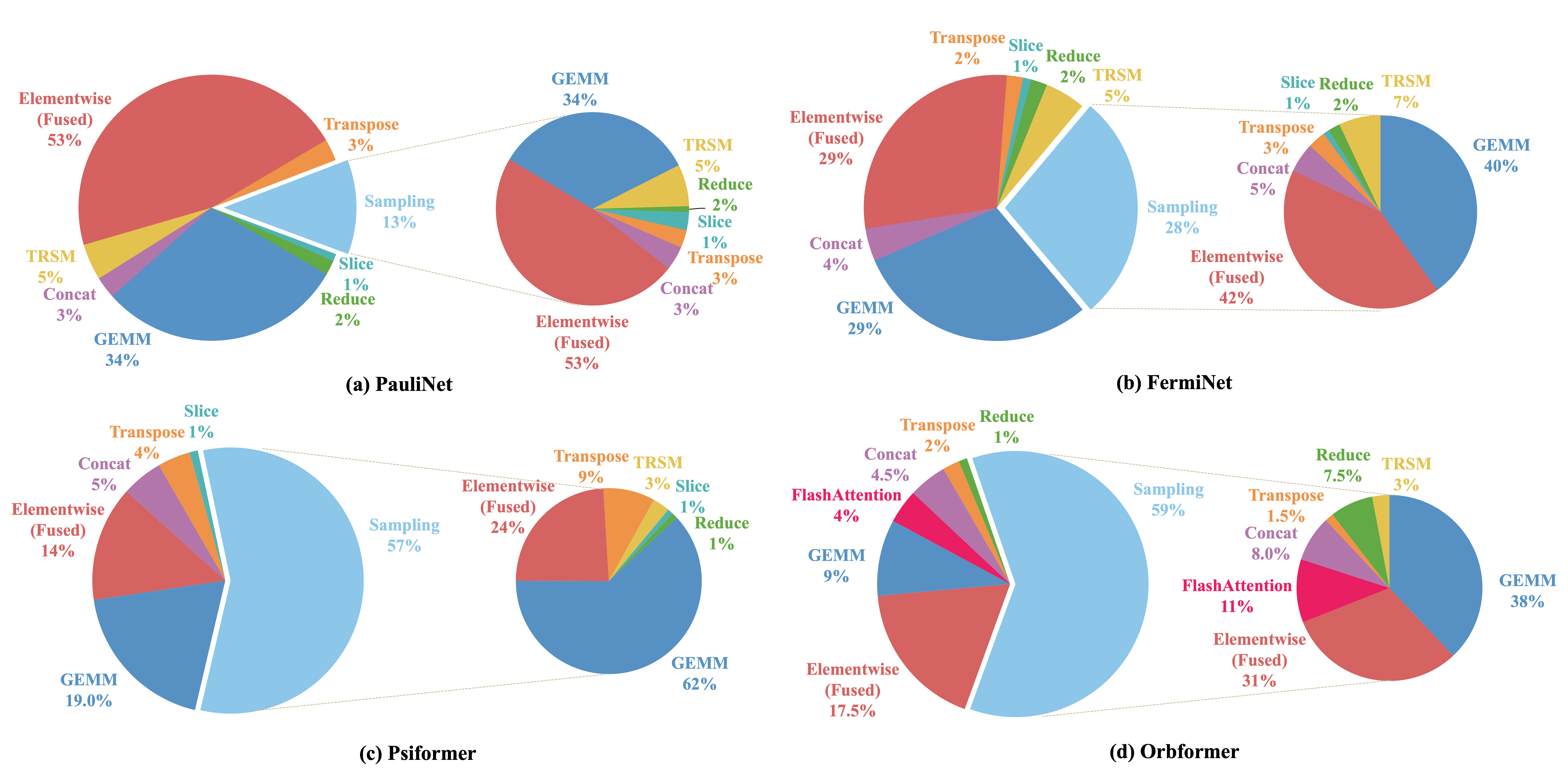}
    \vspace{-2em}
    \caption{Kernel-level runtime breakdown for (a) PauliNet, (b) FermiNet, (c) Psiformer, and (d) Orbformer on an NVIDIA RTX~A5000 GPU.}
    \label{fig:kernel-breakdown}
\end{figure*}

\subsection{Runtime and Memory Summary}
\label{sec:profile-runtime-memory}

Using the unified profiling setup in Section~\ref{sec:profile-settings}, we first examine model-level behavior before moving to kernel-level attribution. Two patterns are most salient: runtime scaling is strongly ans\"atz-dependent, and memory behavior differs across implementation stacks (Figure~\ref{fig:runtime-log}).

\textbf{Observation 1 (runtime scaling and ranking).} Runtime scales with molecule size, but the slope is ans\"atz-dependent. On A5000, kernel time is in the hundreds of milliseconds for LiH/CH$_4$ and reaches seconds for C$_2$H$_6$/C$_4$H$_4$. PauliNet and FermiNet scale steeply from LiH to C$_4$H$_4$ (about $30$--$42\times$ and $25$--$32\times$ across training/inference), while Psiformer and Orbformer scale more mildly ($\sim$8--9$\times$) (Figure~\ref{fig:runtime-log}). The runtime ranking also changes with system size: FermiNet is fastest on LiH and CH$_4$, whereas Psiformer is fastest on C$_2$H$_6$ and C$_4$H$_4$, and PauliNet is slowest at larger systems. The main driver is Laplacian strategy and kernel mix. PauliNet/FermiNet use automatic differentiation (AD) and JVP replay with roughly $3N_{\mathrm{e}}$ re-executions of Stages~A--D, which amplifies fine-grained elementwise and layout kernels. Psiformer/Orbformer reduce this replay overhead with a forward Laplacian path, so more time shifts to sampling and larger GEMM/attention kernels.

\textbf{Observation 2 (memory scaling and hardware gains).} Memory trends also depend on implementation stack. Training memory is consistently higher than inference memory. PauliNet/FermiNet grow relatively smoothly with molecular size, whereas Psiformer/Orbformer on A100 and H200 show step-like jumps near powers of two at C$_2$H$_6$ and C$_4$H$_4$ in the two memory panels of Figure~\ref{fig:runtime-log}. This pattern is consistent with \textsc{oneqmc} allocator effects (dynamic allocation and memory-pool growth granularity), so these points should be interpreted as effective allocated memory rather than pure model-state size. Across hardware, moving from A5000 to A100 yields about $1.1\times$--$2.9\times$ speedup, while H200 yields about $1.7\times$--$5.8\times$, with the largest gains on C$_4$H$_4$. On A5000, training and inference are close for CH$_4$ and larger systems (typically within $\sim$15\%), while LiH shows larger gaps for some ans\"atze.

\takeaway{Runtime by Ans\"atz, Memory by Stack}{At the model level, runtime trends are mainly driven by ans\"atz design and Laplacian strategy, while memory trends are strongly shaped by implementation-level allocation behavior.}

\subsection{PauliNet Workload Characterization}
\label{sec:profile-paulinet}

We use PauliNet as the baseline workload for cross-model comparison.
The following analysis focuses on kernel-family composition and bottlenecks along the Stage~A--E pipeline.

PauliNet combines neural embeddings with Slater determinants, a Jastrow factor, and backflow~\cite{hermann2020deep}, and uses AD/JVP for Laplacian evaluation. This makes it a representative PauliNet workload in the \textsc{deepqmc} implementation for profiling.

\textbf{Observation 1 (kernel composition and phase share).} PauliNet is dominated by fused elementwise kernels, with MCMC at 13\% of total runtime, fused elementwise at 52\%, GEMM at 20\%, and triangular solve matrix (TRSM) at 5\% in the non-sampling slices of Figure~\ref{fig:kernel-breakdown}(a).
\textit{Connection with model architecture:} The Stage~E JVP replay generates many fine-grained elementwise kernels, while Stages~A--D add additional elementwise and layout work around ElectronGNN blocks and wavefunction assembly.

\textbf{Observation 2 (roofline position and utilization).} Most PauliNet kernel families cluster at $10^{-2}$--$10^{-1}$ FLOP/byte, and only GEMM kernels in Stages~A--C approach the compute roof (Figure~\ref{fig:roofline}(a)). PauliNet hotspots remain underutilized, with about 26\% peak instruction throughput, about 14\% fused multiply-add (FMA) activity, around 27\% SM memory throughput, and roughly 48\% level-2 (L2) cache hit rate.
\textit{Connection with model architecture:} The pipeline alternates between GEMM and low-intensity elementwise/layout kernels, and repeated Stage~E derivative paths create small-kernel granularity with limited reuse; end-to-end behavior therefore remains memory-limited.

\takeaway{Stage-E Replay Dominates PauliNet}{PauliNet remains a heterogeneous workload in which fused elementwise and layout kernels dominate total runtime, especially along the Stage~E replay path~\cite{schatzle2023deepqmc}. End-to-end performance is therefore constrained more by memory traffic and kernel granularity than by peak FLOP capacity~\cite{williams2009roofline}.}

\subsection{FermiNet Workload Characterization}
\label{sec:profile-ferminet}
We analyze FermiNet against PauliNet to isolate how ans\"atz design changes runtime composition.
Relative to PauliNet, FermiNet shifts a larger non-sampling share toward GEMM kernels and reduces fused elementwise dominance.

\textbf{Observation 1 (kernel composition and phase share).} FermiNet shifts non-sampling runtime toward GEMM and away from fused elementwise kernels relative to PauliNet, with GEMM increasing from 20\% to 30\%, fused elementwise decreasing from 52\% to 29\%, and MCMC at 28\% (Figure~\ref{fig:kernel-breakdown}(b)).
\textit{Connection with model architecture:} This shift matches FermiNet's higher-capacity ElectronGNN design (more blocks, wider embeddings, and a simpler update head), which increases dense linear-algebra workload while Stage~E still introduces many fine-grained kernels.

\begin{figure*}[!t]
    \centering
    \vspace{-1em}
    \includegraphics[width=\linewidth]{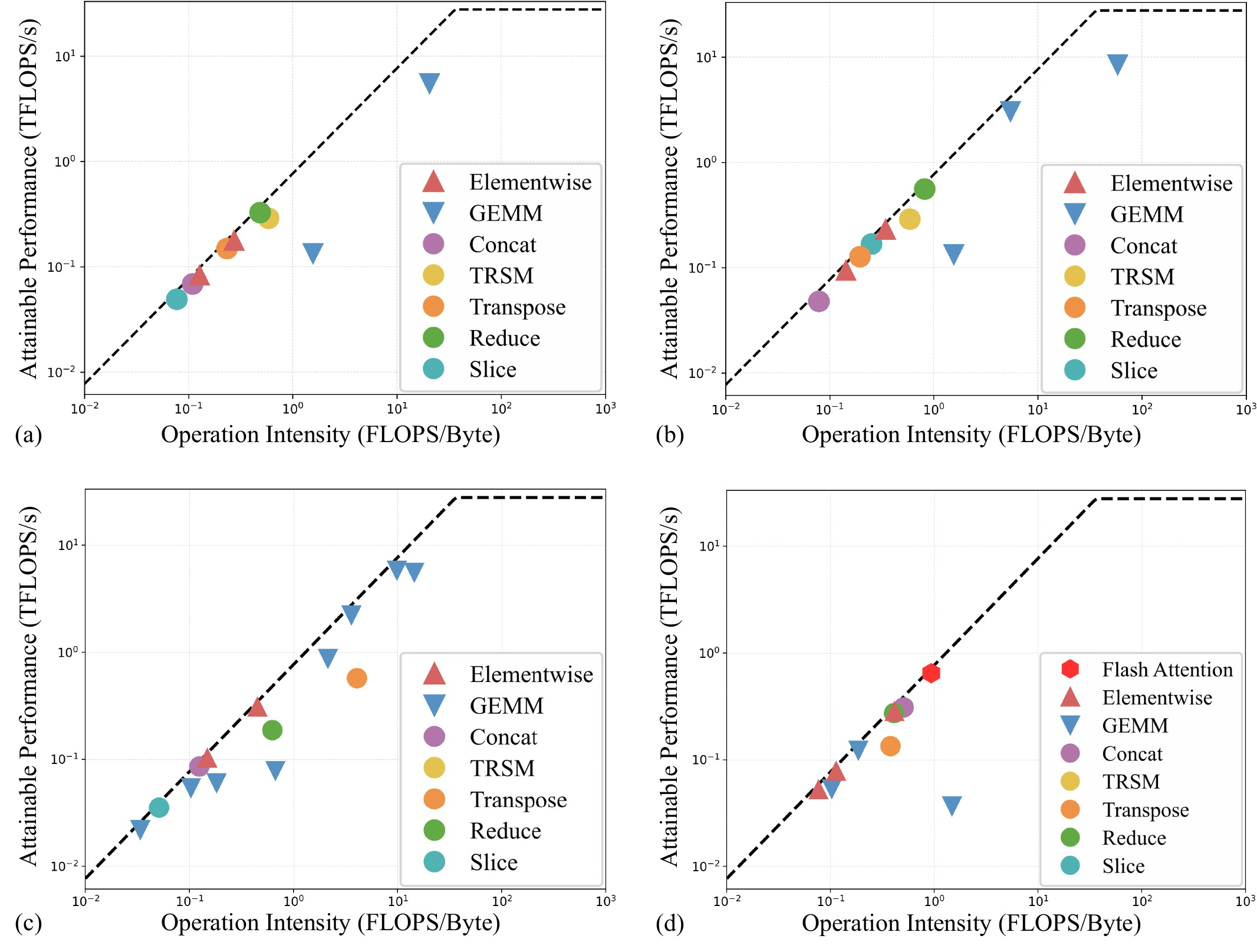}
    \vspace{-2em}
    \caption{Roofline analysis~\cite{williams2009roofline} for (a) PauliNet, (b) FermiNet, (c) Psiformer, and (d) Orbformer on an NVIDIA RTX~A5000 GPU.}
    \label{fig:roofline}
\end{figure*}

\textbf{Observation 2 (roofline position and utilization).} FermiNet places a larger GEMM share in higher-arithmetic-intensity regions, but end-to-end behavior remains memory-sensitive; hotspot counters are still modest, with about 22\% peak instruction throughput and around 50\% L2 hit rate (Figure~\ref{fig:roofline}(b)).
\textit{Connection with model architecture:} Compared with PauliNet, FermiNet relies more on neural blocks and less on explicit physics-inspired terms, which improves GEMM intensity but does not remove bandwidth pressure from Stage~E replay and tensor-layout operations~\cite{hermann2020deep,pfau2020ab}.

\takeaway{More GEMM, Still Memory-Sensitive}{Relative to PauliNet, FermiNet moves more runtime into higher-intensity GEMM kernels as model capacity grows~\cite{schatzle2023deepqmc}. Even so, low-intensity elementwise and layout kernels continue to limit end-to-end utilization through memory traffic and kernel granularity~\cite{williams2009roofline}.}

\subsection{Psiformer Workload Characterization}
\label{sec:profile-psiformer}

Psiformer shifts the workload profile toward compute-intensive GEMM operations and concentrates more runtime in the sampling phase, driven by Transformer-based interactions and a reduced-cost Laplacian path.

\textbf{Observation 1 (kernel composition and phase share).} MCMC sampling accounts for 57\% of end-to-end kernel time. Within sampling, GEMM dominates at 62\%, followed by elementwise (24\%), transpose (9\%), TRSM (3.2\%), and slice/reduce (1\%). In non-sampling phases, GEMM remains the largest component at about 19\%, followed by elementwise at 14\%. Relative to PauliNet/FermiNet, GEMM appears more frequently and at higher arithmetic intensity in the roofline, while transpose shifts upward and slice/reduce shifts toward the memory-bound side.
\textit{Connection with model architecture:} Transformer attention in Stage~B introduces substantial query-key-value (QKV)-related matrix computation, increasing GEMM share and shifting part of execution toward higher-intensity regions.

\textbf{Observation 2 (roofline position and utilization).} Hot-kernel counters increase versus PauliNet/FermiNet: peak instruction throughput rises from about 22\%--26\% to 42\%, SM memory throughput from about 21\% to 31\%, and combined compute/memory throughput from about 34\% to 53\%. Tensor Core utilization remains low on average (about 19\%) but with high variance (up to 76--78\% for a subset of kernels), and L2 hit rate increases slightly to about 60\%.
\textit{Connection with model architecture:} In the \textsc{oneqmc} implementation, the Laplacian strategy in Stage~E reduces replay overhead compared with PauliNet/FermiNet, so a larger fraction of time shifts to sampling-stage kernels.

\takeaway{Sampling-Heavy and GEMM-Heavy}{Compared with PauliNet/FermiNet, Psiformer exhibits a more GEMM-intensive profile with higher sampling-phase concentration, but end-to-end behavior remains heterogeneous and not purely compute-bound.}

\subsection{Orbformer Workload Characterization}
\label{sec:profile-orbformer}

Orbformer shifts the workload back toward a memory-bound regime. FlashAttention and additional MPNN modules reduce GEMM dominance and increase elementwise/data-movement share.

\textbf{Observation 1 (kernel composition and phase share).} Compared with Psiformer, GEMM is roughly halved in both phases (62\% to 31\% in sampling, 19\% to 9.3\% in non-sampling), while elementwise rises (24\% to 38\% in sampling; 14\% to 17.2\% in non-sampling). Sampling increases slightly from 57\% to 59\%. The \texttt{mhsea} FlashAttention family appears with 11\% of sampling time and 3.6\% of non-sampling time; transpose drops from 9\% to 1.5\% in sampling, while concat/reduce increase.
\textit{Connection with model architecture:} FlashAttention fuses previously separate attention-related operations, reducing explicit GEMM/transpose calls, while Orbformer-specific MPNN modules introduce additional elementwise and aggregation kernels.

\textbf{Observation 2 (roofline position and utilization).} Relative to Psiformer, GEMM/transpose/concat kernels shift toward lower arithmetic intensity, and the overall roofline distribution becomes more concentrated in the memory-sensitive region. Hot-kernel counters also decline slightly: top-100 SM utilization is 37\%, SM memory throughput is 29\%, L2 hit rate is 51\%, and combined compute/memory throughput is 51\%.
\textit{Connection with model architecture:} The combined effect of FlashAttention restructuring and added MPNN computation shifts the end-to-end balance away from GEMM dominance toward a more memory-bound operator mix.

\takeaway{Lower GEMM, Stronger Memory-Bound Trend}{Compared with Psiformer, Orbformer exhibits a stronger memory-bound profile, with reduced GEMM share and increased elementwise/data-movement cost.}

\section{Discussion}
\label{sec:discussion}

Section~\ref{sec:profiling} suggests two coupled facts about NNVMC execution. First, many dominant kernels are low-arithmetic-intensity elementwise/layout operators concentrated in the memory-bound roofline region~\cite{williams2009roofline} (Figures~\ref{fig:kernel-breakdown} and~\ref{fig:roofline}), so data movement and kernel granularity can be major bottlenecks. Second, memory-bound dominance does not mean GEMM is negligible: GEMM in embedding/message-passing stages (Stages~A--C) still contributes a substantial runtime share and becomes more prominent in some higher-capacity model variants (e.g., FermiNet vs.\ PauliNet).
This combination indicates that NNVMC can behave as a mixed and phase-dependent workload rather than a single-kernel regime. In particular, Stage~E derivative/Laplacian evaluation replays earlier computations and amplifies fine-grained kernels, which can shift the effective operator balance across stages and ans\"atze~\cite{baydin2018automaticdifferentiation,schatzle2023deepqmc,foster2025ab}. Therefore, compute-only or memory-only tuning may be insufficient.

Based on these observations, this section discusses five algorithm--hardware co-design directions: PIM for memory-bound kernel clusters, collaborative GPU--PIM partitioning, phase-aligned reconfigurable acceleration, architecture support beyond attention kernels, and CPU/SSD memory offloading~\cite{mutlu2020pimprimer,wei2023reconfigurability,krogel2016qmc,he2025waferllm,sheng2023flexgen}. Table~\ref{tab:discussion-map} summarizes how findings in Section~\ref{sec:profiling} map to these directions.

\begin{table*}[!t]
\centering
\caption{Mapping workload findings in Section~\ref{sec:profiling} to the co-design directions discussed in Section~\ref{sec:discussion}.}
\label{tab:discussion-map}
\small
\setlength{\tabcolsep}{5pt}
\renewcommand{\arraystretch}{1.28}
\rowcolors{2}{black!4}{white}
\begin{tabular}{>{\raggedright\arraybackslash}p{0.19\linewidth} >{\raggedright\arraybackslash}p{0.29\linewidth} >{\raggedright\arraybackslash}p{0.20\linewidth} >{\raggedright\arraybackslash}p{0.26\linewidth}}
\toprule
\rowcolor{black!12}
\textbf{Direction} & \textbf{Evidence from Section~\ref{sec:profiling}} & \textbf{Primary Bottleneck} & \textbf{Co-design Opportunity} \\
\midrule
PIM for memory-bound kernels & Elementwise/layout kernels dominate large runtime shares and stay at low arithmetic intensity across models. & Data movement and bandwidth pressure in low-intensity kernels. & Coarse-grained near-memory offloading for memory-bound kernel clusters can be promising when transfer overhead is controlled. \\
\addlinespace[3pt]
Collaborative GPU--PIM system & Execution alternates between GEMM-heavy phases and memory-bound phases, with phase balance varying by ans\"atz. & One fixed substrate mapping may underutilize hardware across phases. & Stage-aware GPU/PIM partitioning with adaptive scheduling can be effective. \\
\addlinespace[3pt]
Reconfigurable acceleration & Operator mix and memory/compute balance shift across stages and model families. & A static compute-memory balance may not fit all phases. & Coarse, phase-aligned reconfiguration modes can help match throughput- vs.\ bandwidth-oriented phases. \\
\addlinespace[3pt]
Architecture support beyond attention & In Orbformer, attention kernels are a limited share of end-to-end runtime, while other kernels remain significant. & Attention-only optimization may yield limited end-to-end gain. & Support for elementwise/layout/data-movement kernels and memory-efficient access patterns may improve end-to-end impact. \\
\addlinespace[3pt]
CPU/SSD memory offload & Memory demand grows non-linearly with system size and can exceed constrained GPU memory budgets. & Capacity bottlenecks in large-system training/inference. & Offloading suitable low-frequency state tensors with asynchronous prefetching can be promising, but requires NNVMC-specific validation. \\
\bottomrule
\end{tabular}
\rowcolors{2}{}{}
\normalsize
\end{table*}

\subsection{PIM for Memory-Bound Kernels}
\discussionsummary{Targeted Challenge: Memory-Bound Kernel Clusters}{Low-intensity elementwise/layout kernels dominate substantial runtime share, so reducing data movement is more impactful than adding peak compute throughput.}

Processing-in-memory (PIM) can move part of computation closer to memory arrays to reduce off-chip data movement and improve bandwidth efficiency for data-movement-dominated kernels.
This direction has been widely explored in top architecture venues for both convolutional/deep neural network (CNN/DNN)-style workloads and Transformer/large language model (LLM) workloads~\cite{kim2016neurocube,chi2016prime,zhou2022transpim,kim2025pimba,lee2025paise}. Fused elementwise and layout kernels operate at very low arithmetic intensity (Figure~\ref{fig:roofline}) and occupy substantial runtime share in PauliNet/FermiNet (Figure~\ref{fig:kernel-breakdown}), so reducing data movement can be more effective than increasing peak compute throughput. PIM is promising for this regime because it executes streaming kernels close to data~\cite{mutlu2020pimprimer,asifuzzaman2023pimsurvey}. In NNVMC, candidate clusters include fused elementwise transforms, reshape/transpose, concatenation, and slicing, which appear in Stages~A--D and become more frequent in Stage~E due to AD/JVP expansion. In practice, PIM may best complement rather than replace GPU execution: GEMM-heavy segments can stay on GPU, while coarse memory-bound clusters can be offloaded to near-memory engines with boundaries/data layouts that amortize transfer and synchronization overhead.

From a methodology perspective, it is also useful to define PIM candidate clusters using trace-level criteria rather than operator names alone. Useful criteria can include low arithmetic intensity, short kernel duration with high invocation count, and repeated layout conversion around stage boundaries. Such criteria can help avoid over-offloading kernels whose memory behavior appears favorable in isolation but whose end-to-end benefit is reduced by migration overhead.

A practical evaluation plan for NNVMC PIM prototypes can report not only runtime speedup, but also energy per optimization step, synchronization overhead ratio, and sensitivity to molecule size. This can provide clearer evidence on when PIM remains beneficial as the Stage~E replay pattern becomes more dominant.

\subsection{Collaborative GPU--PIM Heterogeneous System}
\discussionsummary{Targeted Challenge: Phase-Varying Substrate Mapping}{A fixed GPU-only or PIM-only mapping is not robust; partitioning must track phase-level operator mix while keeping transfer boundaries coarse.}

NNVMC alternates between GEMM-dominated segments and segments with many small memory-bound kernels. In Section~\ref{sec:profiling}, GEMMs are concentrated in embedding and message passing (Stages~A--C), while much of the remaining time is spent in fused elementwise and layout transforms (Figure~\ref{fig:kernel-breakdown}). Stage~E can strengthen this pattern because AD/JVP Laplacian evaluation replays earlier stages and generates many fine-grained kernels. The balance also changes by ans\"atz (e.g., PauliNet vs.\ FermiNet), so a fixed mapping to one substrate may be less robust~\cite{qian2025deep}.

A collaborative GPU--PIM system can keep GEMM-heavy blocks on GPU and offload memory-bound clusters to near-memory execution at coarse granularity~\cite{mutlu2020pimprimer}. Profiling signals from Section~\ref{sec:profiling}, including arithmetic intensity and kernel-family breakdowns, can help choose a small set of cut points (for example near stage boundaries or major blocks) where data layout is already materialized. Inspired by ORCHES~\cite{li2025orches}, one practical strategy can be offline partitioning plus online scheduling: offline, representative traces can be used to choose GPU vs.\ PIM clusters; online, assignment can be adjusted across phases to reduce imbalance and idle time. Here ORCHES is used as scheduling methodology inspiration, not as direct empirical evidence on NNVMC. Transfer and synchronization remain key constraints, and overly fine-grained offloading may remove the benefit even for memory-bound kernels.

For deployment, a lightweight runtime policy can further improve robustness. For example, a scheduler can switch among a small number of pre-profiled mappings based on observed queueing delay and achieved bandwidth during training iterations. This may reduce sensitivity to phase imbalance when the operator mix changes across molecules or ans\"atze.

\subsection{Reconfigurable Acceleration}
\discussionsummary{Targeted Challenge: Static Accelerator Balance}{One static compute--memory design cannot match all phases; practical gains require coarse reconfiguration modes aligned to stage transitions.}

Section~\ref{sec:profiling} suggests that one fixed balance between compute throughput and memory bandwidth may not fit all NNVMC phases or ans\"atze. Even between PauliNet and FermiNet, time excluding sampling shifts between GEMM and fused elementwise kernels, while Stage~E replay often introduces many low-intensity kernels and layout operations. Therefore, an accelerator optimized for one phase (e.g., GEMM-heavy message passing) can be underutilized when execution moves to more memory-bound phases.

Reconfigurable accelerators can address this variability by adapting compute fabric and data paths to the current workload mix~\cite{wei2023reconfigurability}. At processing-element level, bitwidth reconfiguration and sparsity support can change arithmetic cost. At processing-element-array level, reconfigurable dataflow, tiling, and buffering can better match layer shapes and reuse patterns, including multi-engine organizations with heterogeneous dataflows. At system level, chip-level reconfiguration ideas such as hybrid parallelism can support operator-aware partitioning and scheduling when pipelines span multiple devices.

For NNVMC, reconfiguration can be most useful when it is coarse and phase-aligned. Profiling can guide a small set of modes: some favor GEMM throughput in embedding/message passing, others favor bandwidth efficiency for elementwise/layout and AD/JVP replay. Switching may best happen at major stage boundaries. This can complement collaborative GPU--PIM partitioning by keeping utilization higher as operator mix changes across phases and ans\"atze.

Overall, Section~\ref{sec:profiling} suggests two practical requirements for future NNVMC systems: bandwidth efficiency and coarse, phase-aligned specialization.
This also motivates a second-level question: once phase-level mapping is improved, which kernel families still limit end-to-end gains? The next subsection addresses this question by examining why attention-only optimization is often insufficient for Orbformer-like workloads.

\subsection{Architecture Support Beyond Attention}
\discussionsummary{Targeted Challenge: Attention-Only Optimization Ceiling}{Attention kernels are not the sole bottleneck in Orbformer; end-to-end speedup requires support for elementwise/layout/data-movement kernels as well.}

In Orbformer, FlashAttention-related kernels account for a limited share of total runtime (Figure~\ref{fig:kernel-breakdown}(d)). Therefore, optimizing attention alone may be unlikely to deliver large end-to-end gains. In our tests, Orbformer with FlashAttention-1 achieves less than 20\% speedup over the original attention implementation for one MCMC sampling step. FlashAttention-3~\cite{shah2024flashattention3}, with optimizations such as warp specialization and asynchronous warp-group pipelining, already reports high hardware utilization on modern GPUs. This suggests that additional attention-only tuning may provide diminishing end-to-end returns for NNVMC unless other dominant kernel families are also addressed.

The kernel breakdown also shows non-negligible elementwise and transpose time beyond FlashAttention and GEMM. These kernels have low arithmetic intensity and are memory-bound. Their performance can therefore be limited by memory throughput and latency. Prior work (e.g., WaferLLM~\cite{he2025waferllm}) shows that data movement, especially transposition and non-contiguous access, can trigger remote accesses and tiling overheads, which reduce bandwidth utilization. Optimizing memory access patterns, such as shift-based key-value (KV) cache reordering~\cite{he2025waferllm}, can reduce latency and improve these kernels. For NNVMC ans\"atze with many memory-dense operators, architectural support for efficient memory access may be a more practical direction than attention-only optimization.

\subsection{CPU/SSD Memory Offload}
\discussionsummary{Targeted Challenge: GPU Memory Capacity Bottleneck}{Memory growth can exceed constrained GPU memory, but NNVMC-specific offload policies still need rigorous tensor-level validation and scheduling design.}

Section~\ref{sec:profiling} shows that NNVMC memory demand can grow non-linearly with system size and ans\"atz complexity. For larger systems (e.g., C$_4$H$_4$), GPUs with limited memory, such as the 24GB A5000, can encounter out-of-memory. This motivates considering offloading selected state to higher-capacity host memory in central processing unit dynamic random-access memory (CPU DRAM) or secondary storage in solid-state drives (SSDs) to reduce GPU memory pressure.

Prior LLM systems work such as FlexGen~\cite{sheng2023flexgen} suggests that offloading can be effective when tensors have regular access patterns, relatively low access frequency, and tolerance to transfer latency. In NNVMC, potential candidates include parts of sampler and optimizer state and other low-frequency intermediate state. With asynchronous prefetching and coarse transfer scheduling, these tensors may be moved off device while preserving throughput.

However, direct evidence for NNVMC-specific offloading policies is still limited. The key open problem is to identify which tensor groups can be offloaded without destabilizing training or harming time-to-solution. Future work should provide controlled ablation studies and execution plans that jointly optimize offload granularity, prefetch distance, and synchronization overhead.

\section{Related Work}
\label{sec:related-work}

\subsection{Profiling Workloads in Other Emerging AI Applications}
Profiling-oriented system studies have expanded to several emerging artificial intelligence (AI) domains beyond quantum chemistry. In neural rendering,~\cite{li2025unirender} identifies shared operator motifs across heterogeneous rendering pipelines and uses them to motivate unified acceleration abstractions. For real-time 3D neural reconstruction, prior works~\cite{li2025hardwarerace,li2023instant} report that workload scalability is often limited by memory traffic and fine-grained kernel behavior rather than nominal peak throughput. In neuro-symbolic AI, cross-layer profiling has also been used to connect model design choices with hardware efficiency and deployment cost~\cite{wan2024neurosymbolic}.

These efforts establish profiling as a practical methodology for model-system co-design, but they target workloads with different computational structures from NNVMC. NNVMC introduces physics-constrained operators, including Markov chain Monte Carlo sampling, Laplacian evaluation, and antisymmetry-preserving computation, which produce a heterogeneous kernel mix that remains insufficiently compared across model families and hardware platforms.
In this study, we fill this gap by providing a unified stage-aware profiling view across four representative NNVMC ans\"atze (PauliNet, FermiNet, Psiformer, and Orbformer), connecting model-level trends to kernel-level bottlenecks under one protocol.

\subsection{Algorithm-Hardware Co-Design in Foundation Models}
Recent large language model (LLM) co-design studies show that accelerator gains are increasingly tied to model-specific execution patterns. Duplex~\cite{yun2024duplex} co-designs a heterogeneous xPU+Logic-PIM device for mixture-of-experts, grouped query attention, and continuous batching in serving pipelines, where xPU denotes an umbrella term for heterogeneous AI-oriented processing units beyond conventional CPUs and GPUs~\cite{xie2018xpu}. Pimba~\cite{kim2025pimba} and PAISE~\cite{lee2025paise} both target memory-bound inference stages with PIM-centric strategies, but at different levels: Pimba proposes a unified architecture for Transformer and post-Transformer serving, whereas PAISE focuses on GPU--PIM scheduling and kernel-level offloading for attention-dominated decoding. Oaken~\cite{kim2025oaken} further couples key-value cache quantization with dedicated quantization/dequantization hardware and memory management to improve throughput under long-context serving. Compared with these LLM-focused systems, our work targets NNVMC workloads whose bottlenecks are driven by physics-constrained operators and stage-dependent kernel mixes rather than attention- or cache-centric pipelines.
Accordingly, our contribution is not a new accelerator design, but a workload-to-co-design bridge for NNVMC: we identify where memory-centric and phase-aware strategies are likely to help, and map empirical profiling evidence to actionable co-design directions in Section~\ref{sec:discussion}.

\section{Conclusion}
\label{sec:conclusion}

This work provides a workload-oriented study of representative NNVMC ans\"atze from a GPU execution perspective by combining a structured survey and unified profiling of PauliNet, FermiNet, Psiformer, and Orbformer. The results suggest that end-to-end performance is frequently constrained by low-arithmetic-intensity elementwise and data-movement kernels, while runtime composition and compute/memory balance vary across ans\"atze and execution stages, especially through different Laplacian evaluation and sampling paths. These observations indicate that scalable NNVMC acceleration may benefit from phase-aware and memory-centric co-design rather than isolated kernel optimization, and motivate future validation on broader molecular settings and hardware platforms with concrete mechanisms such as heterogeneous partitioning, stage-aware scheduling, and memory offloading.

\bibliographystyle{IEEEtran}
\bibliography{ref}

\end{document}